\documentstyle[12pt]{article}
\newcommand{\be}{\begin{equation}}
\newcommand{\ee}{\end{equation}}

\newcommand{\bn}{\begin{enumerate}}
\newcommand{\en}{\end{enumerate}}

\newcommand{\bea}{\begin{eqnarray}}
\newcommand{\eea}{\end{eqnarray}}
\newcommand{\ba}{\begin{array}}
\newcommand{\ea}{\end{array}}
\newcommand{\lsim}{{\;\raise0.3ex\hbox{$<$\kern-0.75em\raise-1.1ex\hbox{$\sim$}}
\;}}
\newcommand{\gsim}{{\;\raise0.3ex\hbox{$>$\kern-0.75em\raise-1.1ex\hbox{$\sim$}}
\;}}

\newcommand{\CR}{\nonumber \\}

\newcommand{\bnu}{\bar{\nu}}

\newcommand{\cL}{{\cal L}}

\newcommand{\da}{\dagger}
\newcommand{\h}{{1\over2}}

\newcommand{\del}{\partial}

\newcommand{\Tr}{{\rm Tr}}

\newcommand{\de}{\delta}
\newcommand{\DE}{\Delta}
\newcommand{\D}{\delta}
\newcommand{\k}{\kappa}

\newcommand{\La}{\Lambda}
\newcommand{\la}{\lambda_{ \gamma }}
\newcommand{\ka}{\kappa_{ \gamma }}

\newcommand{\ie}{{\it i.e. }}

\newcommand{\ssu}{$SU(2)_L\times SU(2)_R\times U(1)_{B-L}\;$}
\newcommand{\sul}{$SU(2)_L\;$}
\newcommand{\sulu}{$SU(2)_L\times U(1)_Y\;$}
\newcommand{\sur}{$SU(2)_R\;$}
\newcommand{\matr}{\left( \begin{array}}
\newcommand{\ematr}{\end{array} \right)}
\newcommand{\g}{\gamma}

\newcommand{\pool}{\frac{1}{2}}

\newcommand{\ga}{{\gamma}}
\newcommand{\lr}{{left-right symmetric model$\;$}}
\newcommand{\rh}{{right-handed $\;$}}

\newcommand{\sm}{{Standard Model$\;$}}
\newcommand{\ssb}{{spontaneous symmetry breaking $\;$}}
\newcommand{\nlc}{{NLC $\;$}}
\newcommand{\lrp}{{left-right symmetric model.$\;$}}
\newcommand{\smp}{{Standard Model.$\;$}}

\newcommand{\nlcp}{{NLC.$\;$}}

\begin{document}



\section*{Preface }

This Thesis is based on research carried out at the Department of Theoretical
Physics and at the Research Institute for High Energy Physics, University of
Helsinki. The work has
been funded mainly  by the University of Helsinki and  the
Academy of Finland. The financial support from CIMO,  Viro S\"a\"ati\"o,
 Emil Aaltonen Foundation, Jenny and Antti Wihuri Foundation
and  Vilho, Yrj\"o and Kalle
V\"ais\"al\"a Foundation is acknowledged with great appreciation.

I am especially grateful to my supervisor Doc. Jukka Maalampi for his
guidance and expertise. His continuous support and friendship
has made our  collaboration extremely fruitful.

I am indebted to Dr. Katri Huitu for many useful discussions and her help
in my research. Most of the work presented  in this Thesis has
been done in collaboration with her.

I owe much to Prof. Masud Chaichian, Prof. Kari Enqvist, Prof. Paul Hoyer,
Prof. Keijo Kajantie,
Doc. Risto Orava, Prof. Aarre Pietil\"a,
Prof. Matts Roos and Doc. Jorma Tuominiemi for
their support and interest to my research. I have also benefited  from
discussions with Dr. Ricardo Gonzalez Felipe, Dr. Reino Ker\"anen,
 Dr. Mikko Laine, Mr. Rego Ostonen,
Mr. Jari Pennanen, Dr. Constantinos Spartiotis, Mr. Raimo Vuopionper\"a and
Dr. Mikko V\"anttinen.  It has indeed been a pleasure to work
with all  these persons.

I would like to thank all  the people
with whom I have interacted
during my stay in Finland   for their warm hospitality and
I hope they will forgive me for not being able to mention them here.
I ask them to accept my gratitude.

I am greatly indebted to my parents and to Angelika for their
encouragement.

\vspace{1cm}

\noindent
\hspace*{90mm}
\begin{minipage}[t]{48mm}
Helsinki, May 23, 1995 \\

Martti Raidal
\end{minipage}

\newpage

\begin{flushleft}
 {\bf Extended electroweak models and their tests in  future colliders } \\
 Martti Raidal \\
 University of Helsinki, 1995
\end{flushleft}

\section*{Abstract}
In this Thesis some possible
tests of physics beyond the \sm
 in the next generation collider experiments are considered.
The main emphasis is put on the processes which may be  studied  in
the non-conventional $e^-e^-$, $e^-\gamma$ and $\gamma\gamma$ operation
modes of the next linear collider (NLC).

The sensitivity of the NLC for testing the gauge boson self-interactions
through
the reaction $e^-\gamma\to W^-\nu$ is investigated.
New bounds for  precision of measurement
of the $\gamma WW$ coupling parameters $\kappa_{\gamma}$ and $\lambda_{\gamma}$
are derived in the case of  polarized beams,  taking into account
 the recent developments in the \nlc  design.

The minimal version of the
Standard Model does not allow neutrinos to have mass as some recent
astrophysical observations and oscillation  measurements seem to
require. In this work high energy tests of a left-right symmetric extension
of the Standard Model, the $SU(2)_L\times SU(2)_R\times U(1)_{B-L}$ model,
in which  small neutrino masses are generated in a natural way, are
investigated.
The analysis is focussed  on the lepton number violating interactions
of  Majorana neutrinos and  doubly
charged Higgs bosons.
An interesting process, the so-called
inverse double-$ \beta$ decay $ e^-e^-\rightarrow W^-W^-$, is investigated and
its high energy behaviour is discussed.  The process is
found to be useful for clarifying the nature of neutrinos and also
for  studying  the Higgs sector of the model.

In order to avoid the hierarchy problem
 the \lr can be  supersymmetrized.
A distinctive signature of the susy left-right model
is found to  be provided by the
decay of  doubly charged higgsino $ \tilde{\DE}^{--}$.
The production of $ \tilde{\DE}^{--}$ in
 all  collision modes of the \nlc is studied.
 The contribution of doubly charged
higgsino as a virtual state to the selectron pair production
is also estimated.

\newpage

\tableofcontents

\newpage

\section*{List of Papers}

\vspace{0.5cm}

This Thesis consists of an introductory review part, followed by
 four research publications:

\begin{description}

\item{{\bf I}} $ \;\;\;$ M. Raidal,

{\em Tests of gauge boson couplings
in polarized $e^-\gamma$ collisions,}

Nucl. Phys. {\bf B 441} (1995) 49.

\item{{\bf II}} $ \;\;$ P. Helde, K. Huitu, J. Maalampi, M. Raidal,

{\em Gauge boson pair production in electron-electron collisions
with polarized beams,}

Nucl. Phys. {\bf B 437} (1995) 305.

\item{{\bf III}} $ \;$ K. Huitu, J. Maalampi, M. Raidal,

{\em  Supersymmetric left-right model and its test in linear colliders,}

 Nucl. Phys. {\bf B 420} (1994) 449.

\item{{\bf IV}} $ \;$ K. Huitu, J. Maalampi, M. Raidal,

{\em Slepton pair production in supersymmetric left-right model,}

 Phys. Lett. {\bf B 328} (1994) 60.

\end{description}

\newpage

\section{Introduction}

Overhelming amount of  experimental data obtained over the past years
proves that the Standard Model \cite{sm} of particle interactions
has been strikingly successful in terms of  experimental predictions.
 The discovery of  the massive
vector bosons at CERN  \cite{wz} gave a direct evidence for
the correctness of the basic ideas of gauge theories applied in the
Standard Model.
 Furthermore, the precision  analysis of
 LEP electroweak data has enabled physicists
to  constrain strictly those  parameters of the model which are not  directly
measured.
 The recent discovery and mass measurement  of the
 top quark by CDF  and D0  experiments \cite{cdf}
at TEVATRON confirms the results of
LEP precision analysis and  strengthens     the  confidence
in the Standard Model. Based on these experimental results, theoretical
arguments  indicate that the \sm indeed  provides a faithful
characterization of the reality at   energies currently used in experiments
 \cite{rujula}.

Despite of the impressive  success of the Standard Model,
 physicists continue looking for alternative or
  more fundamental theories which
might replace the Standard Model at higher energies.
The large number of free parameters  entering in
the Standard Model,
the maximal parity violation of the weak interaction,
the vanishing neutrino masses and  the hierarchy problem
are among the puzzles which motivate the searches for new models.
On the experimental side, there is still no direct measurement
of one central feature  of  \sul gauge symmetry,
the  non-Abelian self-couplings of $W, Z$ and photon,
and  no direct confirmation
of the Higgs mechanism \cite{higgs}, which  generates
masses in the Standard Model.

A straightforward
way to obtain  hints for   physics beyond the \sm \linebreak
could be the precision study of  gauge boson self-couplings.
Strongly constrained by the gauge invariance, they are
particularly sensitive
to the deviations from the \sm structure.
The new generation particle colliders currently under planning
\cite{lep, lhc, nlc} will give us a possibility to probe
the gauge boson self-couplings in various processes in different
 collision modes.

Another possible approach is to search for  models with extended
gauge symmetries.
One of the most natural extensions of the Standard Model,
the left-right symmetric model, was first proposed
by J.C. Pati and A. Salam in 1974 \cite{pati}.
In addition to the original idea of providing an explanation
to the parity violation of the weak interaction, this model
turned out to be capable of explaining  the lightness of
neutrinos via the so-called see-saw mechanism \cite{seesaw} in a dynamical way.
The observed  solar \cite{solar} and atmospheric \cite{atm} neutrino deficit,
 the recent
$ \overline{\nu}_{ \mu }\rightarrow \overline{\nu}_{ e } $
oscillation events of LSND
experiment \cite{lsnd}, as well as
the COBE satellite hints for possible existence of
a hot component of dark matter
\cite{cobe} seem indeed to indicate
that neutrinos may have small nonvanishing mass.
Although the concept of  the left-right model is already twenty
years old, the  present interest in the model can be explained
by its accessibility  to  experimental tests in the near future.
The energy range of  new colliders could allow us to
study the production of  heavy right-handed
particles, as well as the lepton number violating processes
predicted by the \lrp

However, similarly to the Standard Model, also the \lr
\linebreak suffers from the
hierarchy  problem, \ie the masses of scalar particles
tend to diverge quadratically. This problem can be
cured by  supersymmetrizing  the theory.
It turns out that in the supersymmetric \lr
 one of the doubly charged
Higgs bosons
 should be relatively light \cite{kati} and accessible  in the
future collider experiments.
The mass of its superpartner, the doubly charged
higgsino, is a  free parameter of the model and can also be light.
Therefore, the search for the doubly charged particles will be
of  great interest in  future colliders.

In this Thesis,  the  extensions
of the \sm   described above are considered
and  their phenomenological  implications are studied.
In particular, the main emphasis  is put on  the signatures of
 new physics in the high energy  $ e^+e^-,$ $ e^-e^-,$ $ e^-\g$ and
$ \g\g$ collisions \cite{nlc} possible to be  explored in the next linear
collider (NLC).

The work is organized as follows. A brief summary of
the original publications,  which the Thesis is based on, is given
at the end of this Introduction.

In Section 2  the general parametrization of
the triple boson vertex is  reviewed.
Further
restrictions on the gauge boson self-interactions are discussed and
their implications on the anomalous form factors of
the couplings are considered.

In Section 3   the basic structure of the \lr is described,
 concentrating in particular on the  Higgs sector and on the  symmetry
breaking mechanism.
The importance of the central novel features of the model, \ie
the Majorana nature of neutrinos and the existence of
doubly charged Higgs bosons, is considered
in various physical processes.

In Section 4  the supersymmetric left-right symmetric model,
considered in the  papers this Thesis
is based on,
is introduced.
The particle  contents
of the model is given,  and the predictions  for
 the masses of  Higgs bosons and the susy partners of the  particles
are discussed.

Section 5 is devoted to the experimental tests of the above
mentioned  models.
Reactions with  the most
distinctive experimental signatures in  all the
collision modes of the future linear collider
are considered.

The conclusions are  given in Section 6.

The original papers are appended at the end of the Thesis.

\vspace*{0.5cm}

\subsection*{Summary of the Original Papers}

{\bf Paper I: Tests of gauge boson couplings in
polarized $e^-\gamma$ collisions.}

Single $W$-boson  production  in $e^-\g$ collisions with
polarized beams is investigated. Helicity amplitudes for general
couplings and masses are derived and their properties are discussed.
The results are applied to study the Standard
Model. Updated estimates of the measurement precision of
the photon anomalous coupling parameters $ \ka,$ $ \la$ at the \nlc with
$ \sqrt{s_{ ee }}=500$ GeV are obtained.
Comparison with  earlier studies, done without taking polarization
into account,  shows  a factor of 3 improvement in the measurement
 precision.

\vspace*{0.5cm}

\pagebreak

\noindent {\bf Paper II: Gauge boson pair production in
electron-electron collisions with polarized beams.}

The $W$-boson pair production  in $e^-e^-$ collisions with polarized
beams is investigated. The helicity amplitudes are derived for general
vector - axial vector
couplings and the conditions for a good high-energy behaviour of the
cross-section are given. The results are applied to the heavy vector
boson production in the context of the left-right symmetric model. The
Ward identities and the equivalence theorem are also discussed.

\vspace*{0.5cm}

\noindent {\bf Paper III: Supersymmetric left-right model
and its test in linear colliders.}

 The phenomenological implications of a supersymmetric left-right
model based on $SU(2)_L\times SU(2)_R\times U(1)_{B-L}\,$ gauge symmetry
testable in the next generation linear
colliders are investigated.
In particular, the emphasis is put  on the doubly charged $SU(2)_R$
triplet higgsino $\tilde\Delta$, which is found to have  very
distinguishing  signatures  for experimental
search. Its production rate    in $e^+e^-$, $e^-e^-$, $e^-\gamma$
and $\gamma\gamma$ collisions is estimated and  its subsequent decays are
considered.
These  processes are found to have a clear discovery signature with a very low
background from other processes.

\vspace*{0.5cm}

\noindent {\bf Paper IV: Slepton pair production in
supersymmetric left-right model.}

 The pair production of  sleptons in
electron-positron collisions is investigated in a supersymmetric
left-right model. The cross section is found to be considerably larger
than in the minimal supersymmetric version of the Standard Model
(MSSM) because of larger number  contributing graphs. A novel process is  a
doubly charged higgsino exchange in u-channel, which makes the
angular distribution of the final state particles and the final
state asymmetries  different from those of the MSSM. It also allows
for the flavour non-diagonal final states $\tilde e\tilde\mu$,
$\tilde e\tilde\tau$ and $\tilde \mu\tilde\tau$, forbidden in the
MSSM. These processes  give  indirect information about
neutrino mixings since they depend on the same couplings as the
Majorana mass terms of  right-handed neutrinos.

\newpage

\section{Anomalous Triple  Boson Couplings}

\subsection{ Self-Interactions of Weak Bosons }

The weak gauge  boson sector of the \sm reflects
two fundamental principles:
the  non-Abelian gauge symmetry of the \sm
based on the  gauge group \sulu and
the spontaneous breaking of the gauge symmetry
 which provides the $W$- and $Z$-bosons with their masses.
An immediate  consequence
of the non-Abelian local gauge symmetry is
the existence of gauge boson self-interactions which arise from
the kinetic  Lagrangian of the gauge fields,
\be
\cL_{gauge}=-\pool\Tr(W_{\mu\nu}W^{\mu\nu})-\frac{1}{4}
\Tr(B_{\mu\nu}B^{\mu\nu}),
\label{kin}
\ee
with the field strengths
\be
W_{\mu\nu}= \del_{ \mu }{\bf W}_{ \nu }-\del_{ \nu }{\bf W}_{ \mu } + i
g[{\bf W}_{\mu },{\bf W}_{ \nu }], \;\;
B_{\mu\nu}= \del_{ \mu }B_{ \nu }-\del_{ \nu }B_{ \mu },
\ee
where $ {\bf W}_{ \mu }=\tau_a W^a_{ \mu }/2$ and $ W^a_{ \mu }$ and
$ B_{ \mu }$ denote the non-Abelian and Abelian
field operators of gauge bosons,
respectively.
The gauge principle leads to the universality of the weak coupling constant,
\ie the strength  of the $ W$-boson coupling to fermions is given by  the same
parameter $ g$ which appears in
the three gauge boson as well as in the four gauge boson self-couplings
in the Lagrangian (\ref{kin}).

The Lagrangian (\ref{kin}) describes only the transverse components
of the gauge bosons. The longitudinal components of the massive gauge
bosons would arise from the mass terms of the Lagrangian  which, however,
are not gauge invariant.  The only known possibility to provide the
gauge bosons
with masses, so that the gauge invariance of the
Lagrangian is preserved,
 is  the \ssb mechanism \cite{higgs}.
 In the \sm the linear
realization of the \ssb is used.
There exists a doublet of Higgs scalars in the theory, the
 neutral component of which obtains  a non-vanishing  vacuum
expectation value.
This gives masses to the gauge bosons, and the   three
unphysical Goldstone bosons associated with the broken symmetries
become the longitudinal components of the vector bosons.
If the Higgs boson is very heavy, the weak boson sector becomes
strongly interacting at high energies \cite{strongw}.
The latter possibility is not considered in this work, and it is  assumed
  that
 the doublet Higgs boson is lighter than ${\cal O}(800)$ GeV which ensures
that the gauge boson self-interaction processes remain perturbative.

To date, there has been no direct tests of either the non-Abelian nature nor
the exact realization of \ssb in the electroweak boson system.
The couplings of $ W $-
and $ Z$-bosons to fermions
are tested with a good accuracy, as these interactions can be studied
at the energies of the present experiments at tree level.
The triple and quartic gauge boson vertices, however, enter the low energy
phenomena  only
through  loop corrections and are therefore quite  poorly tested.
The existing $ p\bar{p}$ colliders measure the photon  coupling
to $ W$-boson with errors larger than 100\% \cite{gww}.
The first reaction which will directly test the
non-Abelian gauge structure of the weak bosons is the pair production
of $ W$-bosons in $ e^+e^-$ collisions at LEP 200 \cite{lep}.
This reaction probes
both $ \g WW$ and $ ZWW$ interaction vertices.  At the \nlc
one will also be able to study the reaction $ e^-\g\rightarrow W^-\nu,$
which gives information about the $ \g WW$ coupling alone. This reaction
is investigated in Paper I of this Thesis.

The possible deviations of the gauge boson self-interactions
from their gauge theory form have in general quite remarkable
effects, since they would  spoil the delicate cancellation among the
different amplitudes dictated by the gauge symmetry. This
would mean that the good high energy behaviour of the total
amplitude is lost and  unitarity is violated at high
energies. To preserve  unitarity  some new physical phenomena
beyond the \sm should become effective above a certain  energy scale $ \La\sim
{\cal O}(1)$ TeV.

The effects of  new physics
below $ \La$ can be described by  effective interaction terms of
 light fields in the
Lagrangian. This is analogous to the situation
 in the Fermi theory where the
weak interactions are described
by the  non-renormalizable pointlike
higher order local operators, which lead to the
violation of unitarity at higher energies. Unitarity is preserved
 by introducing a massive vector boson $ W$ to the theory.
In this particular case  the scale $ \La $ can be identified with the
mass of the gauge boson, $ M_W.$
In general,   the parameter $ \La$ should be regarded as
a physical cutoff scale where the  theory has to be replaced by a
more fundamental one.

 There are two different  approaches  to  parametrization
 of the anomalous triple boson interactions.
In the first approach, the so-called
standard phenomenological parametrization \cite{hagi},
one writes down the most general Lagrangian
involving the interaction of three gauge bosons which is allowed by the
Lorentz invariance alone.
Being the most general, this approach can be criticized \cite{rujula, einhorn}
on the basis
that it does not respect the \sulu gauge invariance.
 However, the phenomenological
Lagrangian can be regarded as  a particular parametrization of a gauge
invariant Lagrangian
written down in the unitary gauge
where only the terms which describe the triple boson
coupling are presented \cite{bur}.
 Because of the gauge non-invariant
formulation, the standard phenomenological
parametrization can be used to study the
effects of anomalous couplings at  tree level only.

The gauge invariant formulation of the anomalous couplings
emerges from the effective field theory \cite{efff}  - the theory where
all the  possible higher dimensional operators allowed by the symmetries
 are added to the \sm Lagrangian:
\be
\cL=\cL_{ SM }+\cL_{ effective },
\ee
\be
\cL_{ effective }=\frac{1}{\La^2}\sum \alpha_{ i }^6{\cal O}^6_i +
\frac{1}{\La^4}\sum \alpha_{ i }^8{\cal O}^8_i + ...
\ee
Here $ {\cal O}^n$ denote the operators of dimension $ n$ and
$ \alpha_i$ are the effective coupling constants.
In this theory  all  the appearing
divergences can be absorbed in the renormalization of the coefficients
of the operators \cite{zep}. Therefore, it is applicable also in
the loop calculations.
If the assumption  that
in the first approximation  new physics can be described by the
lowest dimensional  operators alone (dimension 6 in the Standard Model)
 is made, then
 all the parameters employed  by the standard phenomenological
approach are not independent any more \cite{zep}.
In the
standard phenomenological parametrization
this would correspond to imposing
 additional  global symmetries \cite{shild}.

\subsection{ Standard Phenomenological Parametrization of
Triple Boson Coupling}

Let us now consider in more detail the standard phenomenological
parametrization of the triple boson coupling applied in Paper I
to the tree level reaction $ e^-\g\rightarrow \nu W^-.$

On  general grounds,  it has been shown that a charged spin-$ J$ particle
can have $ (6J+1)$  electromagnetic form-factors including
$ C$-, $ P$- and $ CP$-violating terms \cite{boud}. There is also the
equal number of invariant form-factors for the spin-1 coupling
to a charged spin-$J$ particle, provided that the following
conditions hold:
\be
\del_{ \mu }V^{\mu}=0, \;\;\; \del_{ \mu }W^{\mu}=0,
\label{cond}
\ee
where $ V^{\mu}$ denotes either the photon or $ Z^0$ field and
$ W^{\mu}$ stands for the $ W^-$ field.
In this case the most general Lagrangian describing the $ \g WW$ and $ ZWW$
couplings in a Lorentz invariant way can be parametrized
as follows \cite{hagi}:
\bea
\cL_{ VWW }/g_{ VWW } & = & ig^V_1(W_{ \mu\nu }^{\da}W^{\mu}V^{\nu}-
W_{ \mu }^{\da}V_{ \nu }W^{\mu\nu})+i\kappa_V W_{ \mu }^{\da}W_{ \nu }
V^{\mu\nu} \CR
& & +\frac{i\lambda_V}{M_W^2} W_{ \lambda\mu }^{\da}W^{\mu}_{ \nu }
V^{\nu\lambda} - g_4^VW^{\da}_{ \mu }W_{ \nu }(\del^{\mu}V^{\nu} +
\del^{\nu}V^{\mu}) \CR
& & +g_5^V\epsilon^{\mu\nu\rho\sigma}V_{\sigma}(W^{\da}_{ \mu }\del_{ \rho }
W_{ \nu }- W_{ \nu }\del_{ \rho }W_{ \mu }^{\da}) +\pool i\tilde{\kappa}_V
\epsilon^{\mu\nu\rho\sigma}W^{\da}_{ \mu }W_{ \nu }V_{ \rho\sigma } \CR
& &+ \frac{i\tilde{\lambda}_V}{2M^2_W}\epsilon^{\nu\lambda\rho\sigma}
W^{\da}_{ \lambda \mu }W^{\mu}_{ \nu }
V_{ \rho\sigma }.
\label{genlag}
\eea
Here $ W_{ \mu\nu }=\del_{ \mu }W_{ \nu }-\del_{ \nu }W_{ \mu }$ and
$ V_{ \mu\nu }=\del_{ \mu }V_{ \nu }-\del_{ \nu }V_{ \mu }$ denote the Abelian
field strength of $ W^-$ and $ \g/Z^0$ fields, respectively, and
$g^V_1,$ $\kappa_V,$ $\lambda_V,$ $ g_4^V,$ $ g_5^V,$ $\tilde{\kappa}_V $
and $ \tilde{\lambda}_V$ are independent form factors.

The conditions (\ref{cond}) imply
that the scalar part of  massive vector bosons can be neglected.
These  conditions are valid for the on-shell massive vector particles,
as follows from the wave equations of $ W$ and $ Z$. In the case of  off-shell
massive vector bosons the scalar components
 do not contribute  if the vector bosons couple to
massless fermions (this fact follows from the Dirac equation).
For the processes where  light fermion masses are negligible  compared with
the energy involved,
 the Lagrangian (\ref{genlag}) provides indeed the most general
parametrization of the triple boson couplings. This is true for the
 processes like $ e^-e^+\rightarrow W^-W^+$ and
$ e^-\g\rightarrow W^-\nu$ which are of  primary interest in
the future experiments at LEP 200 and \nlcp

In the pair production of the electroweak bosons in $ e^-e^+$
collisions, where both $ \g WW$ and $ ZWW$ couplings play a role,
there are altogether   14 independent parameters.
The reaction $ e^-\g\rightarrow W^-\nu$ studied in Paper I
involves  the $ \g WW$ vertex only, and in the most general
case it depends on 7 parameters.

On the other hand, we know from  experiments that not
all of the parameters involved in the general coupling
can be treated on the same footing. There are well established
symmetry principles which are known to hold with a good
accuracy.  Imposing these  symmetries decreases the number of free
parameters in  the general
Lagrangian (\ref{genlag}). Obviously,
  electromagnetic $ U(1)_{ em }$ gauge invariance is such a symmetry.
It fixes the electric charge of $ W^-$ to be equal to
$ e,$ implying $ g_1^{\g}=1.$
Since in general the $ CP$-violating effects are measured to be  very small,
it is safe to assume that they do not have any detectable impact on
$ \g WW$ coupling. Thus, in the first approximation
the $ CP$-violating terms can
also be neglected.

The most general  Lagrangian in agreement with these assumptions is of the
form
\be
\cL_{ \g WW }  =  -ie( W_{ \mu\nu }^{\da}W^{\mu}A^{\nu}-
W_{ \mu }^{\da}A_{ \nu }W^{\mu\nu}+\ka W_{ \mu }^{\da}W_{ \nu }
F^{\mu\nu}
 +\frac{\la}{M_W^2} W_{ \lambda\mu }^{\da}W^{\mu}_{ \nu }
F^{\nu\lambda}),
\label{lag}
\ee
where $ A_{ \mu }$ and $ F_{ \mu\nu }$ denote the photon field and
the photon field strength, respectively.
 The static electromagnetic properties of
$ W^-$, the magnetic dipole moment $ \mu_W$
and the electric quadrupole moment $ Q_W$, are related  to the
parameters appearing in (\ref{lag}) through  \cite{aro}
\be
\mu_W=\frac{e}{2M_W}(1+\ka+\la),
\ee
\be
Q_W=-\frac{e}{M_W^2}(\ka-\la).
\ee
In the \sm at  tree level one has
$ \ka=1$ and $ \la=0.$
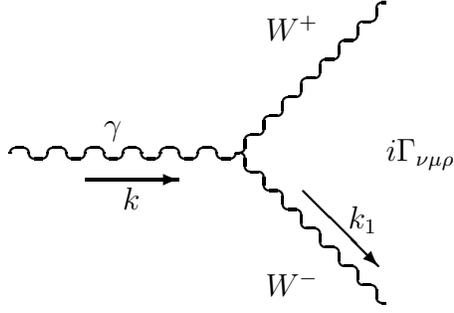
\begin{figure}

\begin{picture}(20,10)(-7,-2)
\thicklines

\multiput(11.5,7.75)(-0.5,-0.5){8}{\oval(0.5,0.5)[br]}
\multiput(12,7.75)(-0.5,-0.5){8}{\oval(0.5,0.5)[tl]}

\multiput(8.5,3.75)(0.5,-0.5){8}{\oval(0.5,0.5)[bl]}
\multiput(8,3.75)(0.5,-0.5){8}{\oval(0.5,0.5)[tr]}

\multiput(2.25,4)(1.,0){6}{\oval(0.5,0.3)[t]}
\multiput(2.75,4)(1.,0){6}{\oval(0.5,0.3)[b]}

\put(4,3.3){\vector(1,0){2.5}}
\put(9.8,3){\vector(1,-1){2}}


\put(8.8,7){\shortstack{\(W^+\)}}
\put(8.8,0.2){\shortstack{\(W^-\)}}
\put(4.5,4.5){\shortstack{\(\gamma\)}}

\put(5,2.5){\shortstack{\(k\)}}
\put(11,2){\shortstack{\(k_{1}\)}}

\put(12,3.8){\shortstack{\(i \Gamma_{ \nu\mu\rho }\)}}

\end{picture}

\caption{Feynman rule for $ \ga WW$ vertex parametrized in
terms of $ \la$ and $ \ka.$}

\label{fivert}
\end{figure}

 In expressing the $ \g WW$ vertex given by the Lagrangian
(\ref{lag}) in the
momentum space  one can take into account
the kinematics of the process at hand.
In the case of the process $ e^-\g\rightarrow W^-\nu$
the  photon and one of the $ W$'s are on mass-shell.
The corresponding vertex has the following form:
\bea
\Gamma_{ \nu\mu\rho }(k, k_1) & = & -2k_{ \nu }g_{ \mu\rho } -
(1+\ka-\la)k_{ 1 \mu }g_{ \nu\rho } + (k+(\ka-\la)k_1)_{ \rho }g_{ \mu\nu }
\CR
& & +\frac{\la}{M^2_W}(k+k_1)_{ \rho }
((k\cdot k_1)g_{ \mu\nu }-  k_{ \nu }k_{1\mu}),
\label{vert}
\eea
where $ k$ and $ k_1$ denote the momenta of the incoming photon and of the
outgoing on mass-shell
$ W^-$, respectively (see Fig.\ref{fivert}).

Are there any general arguments which enable us to estimate the
magnitude of the
 deviations from the Standard Model?   If the underlying
theory is a gauge theory conserving \sulu, then  anomalous
triple boson interactions can occur at the earliest
at one loop level \cite{arzt}.
Since  there is one power of a gauge coupling associated with
every vertex and a loop phase space factor $ 1/16\pi^2$
suppressing    the anomalous coupling, then even in the
case of constructive contribution of several particles
one would not expect any deviations from the \sm couplings
to be larger   than $ \sim 1\%.$
This is roughly  the accuracy which one can expect to be achieved in the
\nlc with c.m. energy $ \sqrt{s}=500$ GeV.
Hence the \nlc will be the first accelerator where one would expect
to achieve  meaningful
constraints on the underlying new physics by studying the anomalous boson
couplings.

\newpage

\section{ Minimal Left-Right Symmetric Model}

One of the still unsolved problems in particle physics
is the understanding of the parity
violation observed in the low energy experiments.
A possible solution is to assume that the interaction Lagrangian
is left-right symmetric, \ie at the level of the Lagrangian
the left- and right-handed fields
are treated on the same basis, but the vacuum is non-invariant under
the parity transformation. That is to say, the left-right symmetry is
spontaneously
broken at very high energies, causing  parity violation effects at
low energies.
This idea is realized  in the  \ssu gauge model \cite{pati}.
At low energies this model reproduces all the features of the
Standard Model, while
at high  energies  parity is conserved and the new particle degrees of
freedom predicted by the model, heavy
gauge bosons, right-handed neutrinos and new Higgs bosons,
start to become manifest.

In addition to a dynamical explanation of the parity violation of
 weak interaction there are several other hints \cite{rabi}
which may indicate that
the left-right symmetry could play a fundamental role in Nature.
One of the most appealing features of the \lr  is that  with
a suitably chosen Higgs sector, it is  capable to explain   the
smallness of  neutrino  masses. The question whether
neutrinos are exactly massless or have a small mass is
not  settled  yet. As mentioned in the Introduction,
there are some astrophysical
observations, \ie  the deficit of solar neutrinos \cite{solar}
and  the possible indication of the COBE
satellite results for a  hot component of dark matter \cite{cobe},
as well as some neutrino oscillation results \cite{lsnd},
which seem to support the idea of massive neutrinos.
In the \lr such  situation can occur in a natural way,
whereas in the \sm the
neutrinos are exactly massless.

Another shortcoming of the \sm is the lack of clear  physical meaning of
the $ U(1)$ generator. In the \lr it becomes the  $ B-L$ quantum number,
the only anomaly free quantum number left ungauged in the \sm, and
all the weak interaction generators have a clear physical meaning.
Thus the Gell-Mann formula for the \ssu theory becomes
\be
Q=I_{ 3L }+I_{ 3R }+\frac{B-L}{2}.
\label{gell}
\ee

While the smallness of $ CP$-violation  is unexplained in the
Standard Model, it is solved dynamically  in the \lr \cite{rabi}.
The suppression of  $ CP$-violating interactions arises due to the
suppression of the  right-handed currents and therefore originates
from the spontaneous breaking  of \ssu symmetry.

Due to the breaking of $ B-L$ symmetry  there exist  lepton
number violating interactions in the \lrp
At very low energies these interactions can be searched for  in
the neutrinoless double-$\beta$ decay experiments \cite{heidel}.
At high energy
collider experiments the $ B-L$ violating interactions
would give rise to   processes with very clean
signatures. One such reaction, the $ W$ pair production
in $ e^-e^-$ collisions, is investigated in Paper II.

Finally, the baryon
number violating interactions of the Majorana neutrinos may
provide us with an explanation of the matter-antimatter asymmetry
in the Universe \cite{buchmyller}.

\subsection{Basic Structure of the  Model}

In the \lr the quark and lepton doublets
\be Q_{ L,R }=\matr{c} u \\ d \ematr_
{ L,R }, \;\;\;
  \psi_{ L,R }=\matr{c} \nu \\ l^- \ematr_{ L,R }
\ee
are assigned to the gauge group \ssu with the  quantum numbers
\be
\left. \begin{array}{llll}
Q_R:\; & (0, \frac{1}{2} ,
\frac{1}{3}), & \;\;\;  Q_L:\; & (\frac{1}{2} , 0, \frac{1}{3}) \\
\psi_R:\; & (0, \frac{1}{2} , -1), & \;\;\;
 \psi_L:\; & (\frac{1}{2} , 0, -1)
\end{array}\right.
\ee
in agreement with eq.(\ref{gell}).
There are  three  coupling constants $ g_R,$ $ g_L$ and $ g'$
associated with the  symmetry groups
\sur, \sul and $ U(1)_{B-L},$ respectively.
For  reasons of  symmetry  the
coupling constants $ g_L$ and $ g_R$ are usually assumed to be equal.
Due to the \sur symmetry the model has two extra gauge bosons,
$ W_R$ and $ Z_R,$ as compared with the Standard Model.
Since there is no experimental evidence for right-handed currents
\cite{mu,klks},
the   new gauge bosons must be much heavier than the ordinary
$ W$ and $ Z$ \cite{tevatron}.

Regarding the Higgs sector of the left-right theories, there are several
possibilities (see \cite{desp} and references therein).
 All the models contain a bidoublet  field
\be
\phi(\pool, \pool, 0)=\matr{cc}
\phi_{ 1 }^0 & \phi_1^+ \\
\phi_2^- & \phi_2^0
\ematr .
\ee
Non-vanishing vacuum expectation
values of its neutral members $ \phi_1^0$ and $ \phi_2^0$ are responsible
for giving  masses to the ordinary gauge bosons
$ W_L$ and $ Z_L$ and contribute also to the masses
of $ W_R$ and $ Z_R.$ They do not break the $ U(1)_{ B-L }$ symmetry,
however.
 In order to
break the $  U(1)_{B-L}$ symmetry, and
to give large enough masses to
the right-handed bosons, one has to add extra Higgs multiplets to the
theory. These additional representations could be, for example,
\sur doublets
with a non-vanishing $  U(1)_{B-L}$ charge \cite{liede},  but in
this case neutrinos would be Dirac particles and
the model fails  to explain the smallness of the neutrino
masses.
In the case of \sur triplet fields the neutrinos are Majorana particles,
and the resulting neutrino mass matrix takes naturally the form
suitable for the see-saw mechanism \cite{seesaw}.
A model with one bidoublet $ \phi(\pool,\pool,0)$ and one
 right-handed Higgs triplet,
\be
 \DE_R(1,0,2)=\matr{cc}
\DE^+_R/\sqrt{2} & \DE^{++}_R \\
\DE_R^0 & -\DE^+_R/\sqrt{2}
\ematr ,
\ee
  is the  minimal scheme, and it  possesses  all the characteristic
features of the left-right symmetric models.

In  \ssb the neutral components of the Higgs multiplets acquire
the vacuum expectation values
\be
<\DE_R>=\matr{cc}
0&0\\
v_R&0
\ematr ,
\ee
\be
<\phi>=\matr{cc}
\k_1&0\\
0&\k_2
\ematr.
\ee
 \ssu symmetry breaks down to  $ U(1)_{ em }$ symmetry
in two stages. In the first stage  the right-handed triplet $ \DE_R$
breaks the initial symmetry \ssu
 to the \sm symmetry \sulu. At the same time
the right-handed charged gauge boson acquires a mass
\be
M_{W_R}^2=\frac{g_R^2v^2_R}{2}.
\ee
By  minimizing
 the triplet Higgs potential
\be
V_{\DE}=-\mu^2\Tr\DE_R\DE_R^{\da}+\rho_1(\Tr\DE_R\DE_R^{\da})^2
+\rho_2\Tr\DE_R\DE_R\Tr\DE_R^{\da}\DE_R^{\da},
\label{tripot}
\ee
one finds  $ v_R=\mu^2/\rho_1.$

The second stage of the spontaneous symmetry breaking,
\ie \sulu $ \rightarrow U(1)_{ em },$
is controlled by the vacuum expectation value of the bidoublet $ <\phi>.$
Apart from giving masses to
$ W_L$ and $ Z_L,$  $ <\phi>$ may slightly mix  $ W_L$ with $ W_R.$
The resulting  mass eigenstates are
\bea
W_1 &=& \cos\xi W_L+\sin\xi W_R, \CR
W_2 &=& -\sin\xi W_L+\cos\xi W_R,
\eea
where the mixing angle $ \xi$ is proportional to $ \xi\sim \k_1\k_2/v^2_R$
\cite{desp}.

During the two-stage  symmetry breaking also the three neutral gauge bosons of
the theory mix, resulting in a physical massless state, the photon,
and two massive states $ Z_1$ and $ Z_2,$ where $ Z_1$ can be identified
with the known neutral weak boson whereas
$ Z_2,$ predominantly the state $ Z_R,$ is heavier.

 Experimental knowledge of the suppression of the right-handed
currents \cite{mu,klks,tevatron} forces us to assume
\be
v_R\gg {\rm max}(\k_1,\k_2).
\label{gg}
\ee
If the right-handed symmetry is broken at a scale much higher than
the electroweak scale, the masses of the right-handed particles,
determined by the new scale, are naturally high. In the gauge boson sector
the mixing between the left and right bosons is
small ($ \xi\rightarrow 0$), and the mass eigenstates $ W_1$ and $ W_2$ with
$ M^2_{W_2}\gg M^2_{W_1}=g^2_L(\k_1^2+\k_2^2)/2$ correspond predominantly to
$ W_L$ and $ W_R,$ respectively.
It has been argued \cite{desp} that in a phenomenologically
consistent model the gauge boson mixing angle should be exactly zero.
In the following we shall always  neglect the
gauge boson mixing and take the weak eigenstates to be equal to the mass
eigenstates. This simplification is supported by the experimental data,
since the mixing angle $ \xi$ is measured to be smaller than $ \xi<0.04$
\cite{pdb}.

Another consequence of the relation (\ref{gg}) concerns the neutrino  masses.
Before studying the  neutrinos of
 \lr explicitly,  let us discuss  possible
neutrino masses in a model independent way.

\subsection{Generation of Neutrino Masses}

The most general mass term for a spin-$ \h$ fermion allowed by
Lorentz  invariance  can be expressed as a sum
$ \cL_{mass}=\cL_D+\cL_M$ of the Dirac mass terms \cite{bil}
\be
\cL_D=m_D\bnu_R\nu_L + h.c.
\label{dir}
\ee
 and the Majorana mass terms
\be
\cL_M=m_L\bar{\nu^c}_R\nu_L + m_R\bnu^c_L\nu_R +h.c.
\label{maj}
\ee
Here $ \nu^c$ denotes the charge conjugated neutrino field defined by
$ \nu^c=C\bnu^T$ where $ C$ is the charge conjugation matrix, and
$ m_D,\; m_L$ and $ m_R$ are constants.
Under the global transformation
\be
\nu\rightarrow e^{i\theta}\nu, \;\;\; \nu^c\rightarrow e^{-i\theta}\nu^c
\ee
the Dirac mass terms $ \cL_D$ transform into themselves,  whereas
the Majorana mass terms $ \cL_M$ acquire an extra phase $ e^{2i\theta}.$
Thus, in contrast to the ordinary Dirac mass terms,
  Majorana mass terms do not
preserve  additive quantum numbers
like  electric charge or  lepton number.
Therefore,  charged fermions can admit only
a Dirac mass  while the neutral fermions like neutrino
can have  mass terms of both types.
If  Majorana  mass terms are present, the theory will contain
interactions which violate  lepton number by two units, $ \DE L=2$.

In the case of a pure Dirac mass term ($ m_L=m_R=0$)
the physical states  are Dirac particles, \ie the neutrino and
the anti-neutrino are different particles.
As can be seen from  the mass Lagrangian (\ref{dir}),
 both left and right chirality
components of a Dirac neutrino have the same mass and
there is no kinematical  suppression of the production of a \rh neutrino.
If both Dirac and Majorana mass terms, or Majorana mass terms alone, are
present, then the physical mass eigenstate neutrinos are their own
antiparticles:
\be
\nu^c=e^{i\lambda}\nu,
\ee
where $ \lambda$ is a real  phase.

Since the mass parameters are arbitrary, one may realize a situation
where $ m_R$ is large compared with $ m_L$ and $ m_D.$
This would ensure that the processes involving
right-handed neutrino interactions are kinematically forbidden
at  low energies.

In the \ssu \lr the neutrino masses arise from the following Yukawa
interaction Lagrangian:
\be
\cL_{ Y }=f\bar{\psi}_L\phi\psi_R + ih\bar{\psi^c}_L\tau_2({\bf \tau\cdot
\DE}_R)\psi_R +h.c. ,
\label{yuk}
\ee
where $ f$ and $ h$ denote unknown Yukawa coupling constants.
The first term in (\ref{yuk}) is of the Dirac type and the second of
the Majorana type.
After  \ssb
neutrinos acquire   masses  which in a suitable basis can be written as
follows:
\be
\cL_{mass }=-(\bar{\nu^c}_R, \bnu_R)
\matr{cc}
0 &  m_D \\
m_D &  m_R
\ematr
\matr{c} \nu_L \\ \nu^c_L \ematr .
\label{seesaw}
\ee
The resulting mass matrix is of the form which  realizes the
see-saw mechanism \cite{seesaw}.
The off-diagonal terms $m_D=f \k_1 $ are determined by the
left-handed symmetry breaking scale and are therefore  expected to be
of the order of a typical Dirac fermion mass. The non-zero diagonal
entry $ m_R=h v_R$ is set by the  breaking scale of the \sur symmetry,
which justifies  the assumption $ m_D\ll m_R.$

The physics content of the Lagrangian (\ref{seesaw})
becomes apparent when it is diagonalized.
It  can be written in the canonical form
\be
\cL_{ mass }=-(m_1\bar{\chi}_1\chi_1+m_2\bar{\chi}_2\chi_2),
\ee
where the eigenvectors $ \chi_1$ and $ \chi_2$
are expressed in terms of the self-conjugate  fields
$\chi_L =\nu_L +\nu^c_R$ and
$\chi_R=\nu_R+\nu^c_L $
as follows:
\bea
\chi_1&\approx&\chi_L-\frac{m_D}{m_R}\chi_R, \CR
\chi_2&\approx&\chi_R+\frac{m_D}{m_R}\chi_L.
\label{eig}
\eea

The masses of these two Majorana neutrinos $ \chi_1$ and $ \chi_2$
are approximately
\be
m_1  \approx  \frac{m_D^2}{m_R}, \;\;\;
m_2  \approx  m_R,
\label{massid}
\ee
and the states  have opposite $ CP$-parities,
\be
\eta_{ CP }(\chi_1)=-i, \;\;\; \eta_{ CP }(\chi_2)=+i.
\ee
 The Majorana nature of the neutrinos is obvious
since $ \nu^c_{L,R}=(\nu_{ R,L })^c.$
 According to eq.(\ref{eig}), the heavy mass eigenstate $ \chi_2$  corresponds
 predominantly to the right-handed neutrino, and the light mass
eigenstate $ \chi_1$
predominantly to the left-handed neutrino. Neglecting the
small mixing of the order of $ m_D/m_R$, the heavy neutrino
interacts via $ (V+A)$ currents, while the light,  Standard-Model-like
neutrino, interacts via $ (V-A)$ currents.
As can be seen from (\ref{massid}), the mass of the light neutrino
is directly connected to the scale $ v_R$ where the left-right symmetry
is broken so that the larger  this scale, the smaller  the neutrino mass.
In the limit
$ v_R\rightarrow\infty ,$
 light neutrino mass vanishes and the \lr becomes in all respects
indistinguishable from the \smp

Despite of the searches for the new gauge bosons and $ (V+A)$
currents in  accelerators as well as in low-energy weak
interaction experiments, no indications of their existence
has been found so far. An interesting possibility for the
future collider experiments would be to investigate the $ |\DE L|=2$
interactions, which in the framework  of the \lr are mediated by
 massive neutrinos and  doubly charged Higgs bosons.
Such  studies will not only reveal the nature of the massive neutrinos,
but also shed light on the Higgs sector of the theory.
In Paper II, one analysis of this type is carried out.

\newpage

\section{Supersymmetric Left-Right Model}

The  \lr has enabled us to explain successfully
 some of the questions left unanswered in the \smp
On the other hand, like in  the Standard Model,
 the \lr suffers from the hierarchy
problem: the masses of the Higgs scalars diverge quadratically
when  loop corrections are taken into account.
One  can cure this problem by making the theory  supersymmetric.
Due to the symmetry between fermions and bosons, there exist
 supersymmetric partners for every ordinary particle,
which  cancel the quadratic divergences.
While there is no evidence for the existence of the susy partners,
the interest in supersymmetry  was renewed when it turned out that
in $ SU(5)$ grand unified model the unification of coupling constants
at high energies occurs only when the model is supersymmetric \cite{lan}.
However, in the $ SO(10)$ theory, which would be the grand unified framework
for the left-right model, supersymmetry would not be necessary  for this reason
\cite{soten}.

Apart from the existence of superpartners,
the biggest difference between susy and non-susy models concerns the
Higgs sector. From the phenomenological point of view
a major difference is that in   susy models some of the Higgs bosons must
be rather light.
It has been shown that in
supersymmetric  models with arbitrary Higgs sector  the mass of
the  neutral \sm like Higgs particle   cannot be heavier than about
$  150$ GeV
\cite{kane}.

Susy \lr has been previously studied by many authors
 \cite{slr1,slr2,frank1,frank5,slr9}.
 Paper III of this Thesis introduces
 a minimal susy left-right model where the number of Higgs fields
is the smallest possible.
The minimal set of Higgs fields in the non-susy left-right model
consists of a bidoublet $ \phi_u$
and a $ SU(2)_R $ triplet $ \DE_R$.
After supersymmetrization, the cancellation of chiral anomalies among the
fermionic partners of the triplet Higgs fields $\DE_R$ requires
 introduction of a second triplet $\de_R$ with opposite $U(1)_{B-L}$
quantum number. Due to  conservation of the $B-L$ symmetry, $\de_R$
does not couple to leptons or quarks. In order to avoid a trivial
Kobayashi-Maskawa matrix for quarks, also another bidoublet $ \phi_d$
 should be
added to the model. This is because supersymmetry forbids a Yukawa
coupling where the bidoublet appears as a conjugate.

The vacuum expectation values for the Higgses, which
break the  $SU(2)_L\times SU(2)_R\times U(1)_{B-L}$ into the
$U(1)_{em}$,  can be chosen as follows:
 \be <\phi_u >=\left( {\begin{array}{cc} \kappa_u & 0\\ 0 & 0
\end{array}}  \right), \: <\phi_d >=\left( {\begin{array}{cc} 0 & 0\\
0 & \kappa_d \end{array}}  \right),\:   <\Delta_R>=\left(
{\begin{array}{cc} 0 & 0\\ v_R & 0 \end{array}} \right)
 ,\:   <\delta_R >\equiv 0. \label{vevs} \ee
\noindent Here  $\kappa_{u,d}$ are of the order of the  electroweak
scale $10^2$ GeV, since they give mass to $ W_L$ and $ Z_L.$
As discussed in the previous Section,
the vacuum expectation  value  $v_R$ of the triplet Higgs has to be
large in order to have the masses of the new gauge bosons $W_R$ and
$Z_R$ sufficiently high.
Since in (\ref{vevs}) only one of the neutral fields in each of the
bidoublets $ \phi_u$ and $ \phi_d$ are assumed to
acquire a non-vanishing  vacuum expectation  value,
the charged gauge bosons do not mix, and $W_L$ corresponds to the
observed  weak bosons.

In Papers III and IV the supersymmetric version of the \lr
and its phenomenological implications for high energy collisions are
investigated.
There it is assumed that  the superpotential has the following form:
 \bea W & = & h_u^Q \widehat Q_L^{cT} \widehat \phi_u  \widehat Q_R  +
h_d^Q \widehat Q_L^{cT} \widehat \phi_d  \widehat Q_R \nonumber \\
&&+h_u^L \widehat L_L^{cT} \widehat \phi_u  \widehat L_R  +h_d^L
\widehat L_L^{cT} \widehat \phi_d  \widehat L_R    +h_\DE \widehat
L_R^{T} i\tau_2 \widehat \DE_R  \widehat L_R \nonumber\\ && + \mu_1
{\rm Tr} (\tau_2 \widehat \phi_u^T \tau_2 \widehat \phi_d )  +\mu_2
{\rm Tr} (\widehat \DE_R \widehat \D_R ) .
\label{pot}
\eea

 \noindent Here $\widehat Q_{L(R)}$ stands for the
doublet of left(right)-handed quark superfields, $\widehat L_{L(R)}$
stands for the doublet of left(right)-handed lepton superfields,
$\widehat \phi_u$ and $\widehat \phi_d$ are the two bidoublet Higgs
superfields, and $\widehat \DE_R$ and $ \widehat \D_R$ the two triplet
Higgs superfields.
In  superpotential (\ref{pot}) the $R$-parity,
$R=(-1)^{3(B-L)+2S}$, is preserved. This and the assumption $ <\tilde{\nu}>=0$
ensures that  susy
partners with $R=-1$ are produced in pairs and that the lightest
supersymmetric particle  is stable.

If  supersymmetry were an  exact symmetry,  the masses of
 particles and their susy partners  would be the same.
Since none of the superpartners has been  observed, one has to
break supersymmetry. In Papers III and IV breaking has been assumed to
happen softly, \ie  the superpotential contains
the most general non-supersymmetric mass terms for
 scalars and gauginos, which do not give rise to  quadratic divergences:
\be
\cL_{ soft }=-\h \sum_i m_i^2|\varphi_i|^2 -\h\sum_{ \alpha }
M_{ \alpha }\lambda_{ \alpha }\lambda_{ \alpha } +B \varphi^2+
A\varphi^3 + h.c.
\ee
Here $ \varphi_i$ denote  scalar fields
and $ \lambda_{\alpha}$ stands for gaugino fields.
 The soft masses should not be much heavier
than ${\cal O} (1)$ TeV \cite{haber,colour}
because their contribution to the lightest Higgs mass
would become too large otherwise.
In
order  to preserve the naturalness of the  theory the
 supersymmetric mass parameters $\mu_i $ in the superpotential
(\ref{pot})  should be
close to the  scale of  soft supersymmetry breaking.
In order to avoid an unnatural hierarchy of the mass parameters
at the Lagrangian level, one can assume  that the parameters
$|\mu_i|$ are also of the order of  the electroweak scale.

Particularly interesting objects from the phenomenological point of view
are  the doubly charged
fermions,  the superpartners of the  triplet scalars $ \tilde{\DE}^{++}$ and
$ \tilde{\D}^{++}.$ Their mass
matrix
is particularly simple, since doubly charged higgsinos do not
mix with gauginos.
The susy mass terms for triplet higgsinos are given by
\be
\cL_{ triplet mass }=-\mu_2 (\tilde{\DE}^+\tilde{\D}^- +
\tilde{\DE}^{++}\tilde{\D}^{--}+ \tilde{\DE}^0\tilde{\D}^0) + h.c.
\ee
which implies that the mass of the doubly charged higgsino is
 set by the parameter $ \mu_2=M_{ \DE^{++} }.$ Thus the
doubly charged higgsino mass is a free parameter of the
model which by  naturality arguments should be close to the weak
scale.
  The triplet higgsinos, like the triplet
 Higgses, carry two units of lepton number,  and therefore
 the final state of their  decay must also have
even lepton number in the case of $R$-parity conservation.
This follows from the
Lagrangian
\be
\cL_{ \tilde{\DE}\tilde{l}l }=-2h_{ \DE }\bar{l^c}\tilde{\DE}\tilde{l},
\ee
which describes an  interaction between the doubly charged
higgsino, lepton and the slepton.

 There are five charginos
$ \psi^{\pm}_j$
and nine neutralinos
$\psi^{0}_i$
in this model.
The physical particles $ \tilde\chi^{\pm}_i$
and $\tilde\chi^{0}_i $ are  found by  diagonalizing   the mass
Lagrangian:
\be \tilde\chi^{\pm}_i=\sum_{j}C_{ij}^{\pm}\psi_j^{\pm},
\label{charginos} \ee
\be \tilde\chi^{0}_i=\sum_jN_{ij}\psi^0_j, \label{chargino} \ee
where $ C_{ ij }^{\pm}$ and $ N_{ ij }$ denote the diagonalizing matrices
of charginos and neutralinos, respectively.
  The neutralinos are
Majorana particles, whereas the charginos combine   to
form Dirac fermions.

 The
masses of susy particles depend on the following parameters: the soft gaugino
masses, the supersymmetric Higgs masses $\mu_1$ and
$\mu_2$, the vacuum expectation values $\kappa_u$, $\kappa_d$, and
$v_R$, and the gauge coupling constants.
It has been
shown \cite{kati} that  one of the doubly
charged Higgses $ \DE^{++}$ in the supersymmetric
left-right model must be lighter than a few
hundred GeV.
In Papers III and IV we have carried out the analysis of  the particle
contents of the model for different values of the mass parameters.
For  large soft gaugino masses (around 1 TeV)
the neutralinos are predominantly
higgsinos, whereas for  smaller soft masses (  200 GeV)
they are mainly gauginos.
The difference in the neutralino   interactions  in these two cases
should manifest itself in  collider experiments.
 For simplicity, in Paper IV the mixing of the left and right selectrons
is  assumed  to be negligible,    and  their masses
$m_{\tilde e_L}$ and
$m_{\tilde e_R}$ are taken  to be equal.

\newpage

\section{Tests of the Extended Models}

The main emphasis in the  Papers of this Thesis  is on
the high energy signatures of the beyond-the-Standard-Model schemes
described in the foregoing Sections. The experimental environment one has in
mind is mainly the \nlcp  According to  present plans \cite{nlc},
in the first stage the \nlc will operate at the $ e^+e^-$ c.m. energy of
$ \sqrt{s}=0.5$ TeV, and later the energy will be increased up to
$ \sqrt{s}=2 $ TeV. The anticipated luminosity of the $ e^+e^-$
collision mode is $ \cL=10^{34}$ cm$ ^{-2}$s$ ^{-1}$.

While the $ e^-e^+$ option will be the
main operation mode of the next linear collider,
also  $ e^-e^-,$ $ e^-\g$ and
$ \g\g$ collisions are technically feasible \cite{nlc}.
The $ e^-e^-$ collider design is, in principle, simpler than the $ e^-e^+$
one since there is no need to create positrons.
For the photon colliders the  photon beam  can be obtained by
scattering linearly polarized laser light off  of the electron beam.
The result is  a polarized photon beam with very
hard spectrum strongly peaked at
the maximum energy which is about $ 84$\% of the electron beam energy
\cite{ginz}.
The growing interest in
 the non-conventional collision modes arises  not only from the need of
having  complementary tests of physical quantities but also
from the fact that for many purposes the new options provide
more useful  reactions  to study  than the conventional
$ e^-e^+$ mode. For example, the $ e^-e^-$ option is particularly suitable
for the study of  lepton number violating interactions, since its initial
state carries lepton number two. The $ e^-\g$ and $ \g\g$ modes
give us a direct access to
processes which in the other collision modes appear
only as  subprocesses of higher order reactions.    Hence
the new collision options of the \nlc are considered  to
be very useful  for  studying various extensions of the \smp

\subsection{Measurements of Anomalous  Triple Boson Coupling}

At   present the most stringent
experimental bounds for the $ \g WW$ vertex parameters
$ \ka$ and $ \la$ defined in the Lagrangian
(\ref{lag}) are  obtained   by the
TEVATRON CDF experiment
from  a direct
measurement of  photon-$ W$ interaction  \cite{gww}:
\[ -2.3\leq 1-\ka\leq 2.2, \]
\be -0.7\leq\la\leq 0.7, \label{cdflimw} \ee
given  at  95\% CL.
More stringent, but less direct
limits have been obtained from  studies of  electroweak
radiative corrections of  low energy data by applying the effective
Lagrangian formalism of  dimension-6 operators using,  however,
 some  additional assumptions  \cite{rujula}.
 It has been
shown in ref.\cite{zep} that there is no rigorous model-independent bounds
for the anomalous triple boson coupling from radiative corrections.
Therefore, the precise
direct measurements are needed in the future colliders
in order to improve the limits (\ref{cdflimw}).

The first electron collider which will probe the $\g WW$ coupling is LEP 200.
Because of the low energy and low luminosity,  LEP 200  will be able
to measure the coupling only with a precision not better than about
10\% \cite{miya}. Theoretically,
one should not expect anomalous couplings to be larger than 1\% \cite{arzt},
and hence  the first stage of the
\nlc  is likely to be the first place  to produce  relevant  new constraints.

The $ W$-boson pair production \cite{hagi}
\be
e^-e^+\rightarrow W^-W^+
\label{ww}
\ee
is a particularly clean  process to study the triple boson couplings  and
to find constraints on the anomalous terms.
It will be an important reaction to be experimentally  studied in the \nlcp
Taking the collision energy to be
  $ \sqrt{s}=500$ GeV,  integrated luminosity
50 fb$^{-1}$ and beam polarization  90\%,
the sensitivity of the \nlc with respect to the parameters $ \ka$ and $ \la$
  at 95\% CL is anticipated to be \cite{miya}
\[ -0.0052\leq 1-\ka\leq 0.0057, \]
\be -0.012\leq \la \leq 0.021 .\label{wwlim}\ee

A  disadvantage of the processes (\ref{ww}) is that
it does  not allow separate tests of the
anomalous photon and $ Z^0$ couplings, since
both $ \g WW$ and $ ZWW$ vertices are involved in the reaction.
The $ e^-\g$ collision
option of  the \nlc will be an ideal place for
studying the photon anomalous couplings separately.

There are two different possible collision schemes for the photon colliders
 \cite{telnov}.  First, the photon conversion region is very
close to the interaction point and  the
entire photon spectrum interacts with the
electron beam. From the physics point of view this realization of the
$ e^-\g$ collisions is undesired because of the high rate
of  background processes initiated
by the electrons which have been  used for creating the photon beam, and
also because of the low monochromaticity of the photon beam.

In  the second collision scheme the
distance  between the conversion and interaction points is larger.
The electrons used for producing the photon beam are removed
by applying a strong magnetic field,
 and therefore the  $ e^-\g$ collisions are  clean and
highly monochromatic.
 The achievable luminosities
in this case are found to vary from 30 fb$^{-1}$ at VLEPP to 200 fb$^{-1}$
at TESLA per year \cite{telnov}
depending on the linear collider design.

The most sensitive process to the photon anomalous coupling
in $ e^-\g$ collision mode is
\be
e^-\g \rightarrow W^-\nu.
\label{proc}
\ee
 The  process
has  been previously investigated in ref.\cite{eg}.
In Paper I  the updated   analysis
of the reaction (\ref{proc}),
taking into account   beam  polarization
 as well as  recent developments in the linear collider
design, is carried  out.
 The center of mass energy is assumed to be
$ \sqrt{s_{ e\g }}=420$ GeV,
corresponding to the peak value of the photon energy spectrum.
The integrated luminosity is estimated to be $ \cL_{ int }=50$ fb$^{-1}.$
A $ \chi^2$ analysis of the differential cross section
of the process, which turns  out to be the most sensitive observable
with regards to the parameters $ \ka$ and $ \la,$ yields  for the
measurement sensitivity
of the $ e^-\g$ collider at 90\% CL the
following bounds:
\[ -0.01\leq 1-\ka\leq 0.01, \]
\be -0.012\leq\la\leq 0.007 .\label{xxx}\ee
The estimate  of Paper I shows  that the beam polarization together
with the monochromaticity of the photon beam improves   the
sensitivity by a factor of $3.$
Comparison with the bounds (\ref{wwlim}) reveals that the process (\ref{proc})
allows  to constrain the parameter $ \la$ more strictly  than
the process (\ref{ww}). Let us emphasize again, the bounds (\ref{xxx})
are independent of $ ZWW$ coupling.

In order to further increase
the measurement precision of the anomalous triple
boson coupling one needs higher collision energies. It has been estimated
in ref.\cite{miya} that
the \nlc with  a center of mass energy $ \sqrt{s}=1$ TeV
will allow measurements with the precision of about 0.1\%.

\subsection{ Searches for  Left-Right Symmetry}

Up to date no new heavy gauge bosons,  heavy neutrinos or
new types of interactions going beyond the \sm have been discovered.
In order to make the \lr consistent with this fact
 one must  constrain the parameters of the model,
such as the masses of the right-handed gauge bosons $ W_R$ and $ Z_R$
and the Lorentz structure of the interactions.

The experimental constraints  are obtained mainly from the low-energy
weak interaction processes,
where one searches for  possible manifestations of $ (V+A)$
currents, and from  high energy collider experiments,
which give direct mass limits for the new particles.
The present lower limit for the  mass of  $ W_R$, obtained
from a direct search  in TEVATRON,
 is $ M_{W_R}\geq 652$ GeV \cite{tevatron}.
This is comparable with the  limits from the low energy
processes $ K\rightarrow \pi\mu\bnu,$ $ \mu\rightarrow e\nu\bnu$
\cite{mu} where $ W_R$ appears as a virtual intermediate state.
 The most stringent limit quoted in the literature
is derived from the
$K_L-K_S$ mass difference \cite{klks}:  $ M_{ W_R }\geq 1.6$ TeV.

All these constraints, however, are subject to various assumptions
about the details of the \lrp
The gauge couplings of
$ SU(2)_{ L }$ and $ SU(2)_{ R }$
gauge groups are usually  taken to be equal $ g_L=g_R$,
the  Cabibbo-Kobayashi-Maskawa matrix for the right-handed quarks
is assumed to be the
same as for the left-handed quarks, $ V_R=V_L$, and the right-handed neutrinos
are assumed to be light.
All the  limits discussed will be considerably
weakened  if these simplifying assumptions are relaxed \cite{sank}.
For example, the muon decay
$ \mu\rightarrow e\nu\bnu$  bounds do not hold if the
mass of the \rh neutrino  exceeds  50 MeV. The TEVATRON bounds
on  $ M_{ W_R }$ are degraded
if the \rh neutrinos decay in the detector, or if $ (V_R)_{ ud }\ll
(V_L)_{ ud }.$ Also the $ K_L-K_S$ limit can be weakened if
$ V_R \neq V_L$ and $ g_R < g_L.$

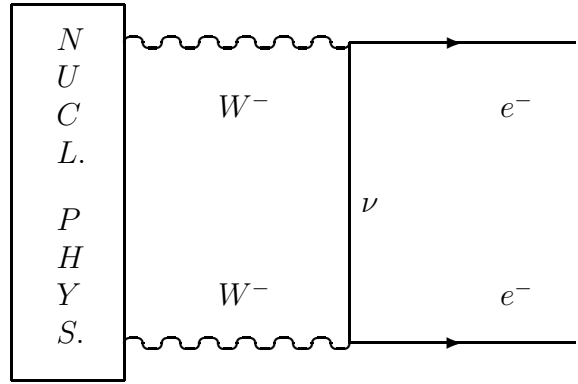
\begin{figure}

\begin{picture}(28,11)(-11,-2)
\thicklines

\put(6,8){\line(1,0){6}}
\put(6,0){\line(1,0){6}}
\put(6,8){\line(0,-1){8}}

\put(6,8){\vector(1,0){3}}
\put(6,0){\vector(1,0){3}}

\multiput(0.25,0)(1,0){6}{\oval(0.5,0.3)[t]}
\multiput(0.75,0)(1,0){6}{\oval(0.5,0.3)[b]}
\multiput(0.25,8)(1,0){6}{\oval(0.5,0.3)[t]}
\multiput(0.75,8)(1,0){6}{\oval(0.5,0.3)[b]}


\put(0,9){\line(0,-1){10}}
\put(-3,9){\line(0,-1){10}}
\put(-3,9){\line(1,0){3}}
\put(-3,-1){\line(1,0){3}}

\put(2.5,1){\shortstack{\(W^-\)}}
\put(2.5,6){\shortstack{\(W^-\)}}
\put(10,6){\shortstack{\(e^{-}\)}}
\put(10,1){\shortstack{\(e^{-}\)}}
\put(6.3,3.5){\shortstack{\(\nu \)}}

\put(-1.8,7.8){\shortstack{\( N\)}}
\put(-1.8,6.8){\shortstack{\( U\)}}
\put(-1.8,5.8){\shortstack{\( C\)}}
\put(-1.8,4.8){\shortstack{\( L.\)}}
\put(-1.8,3){\shortstack{\( P\)}}
\put(-1.8,2){\shortstack{\( H\)}}
\put(-1.8,1){\shortstack{\( Y\)}}
\put(-1.8,0){\shortstack{\( S.\)}}

\end{picture}

\caption{Feynman diagram giving rise to the neutrinoless
double-$ \beta$ decay.}

\end{figure}

Experimentally the  most intriguing  prediction of the \lr
is the existence of  lepton number violating processes associated with
the Majorana nature of neutrinos.
One of the processes studied  at low energies is the neutrinoless
double-$ \beta$ decay \cite{heidel},
\be
(A,Z)\rightarrow (A,Z+2)+2e^-,
\label{double}
\ee
where a nucleus decays into another nucleus by  emitting  of two electrons.
In the \lr this reaction  can arise from the diagram depicted in Fig. 2
where the gauge bosons   can  be either an ordinary $ W_L$
or a \rh $ W_R.$ The important feature of the process is that it takes
place only when the mediated neutrino is a massive Majorana particle.
In the \sm the lepton number is conserved and the reaction (\ref{double})
is forbidden.
So far, there exists no evidence for the neutrinoless double-$ \beta$
decay. This can be used to derive  constraints for the light neutrino mass
which at the moment is $ \langle m_{ \nu }\rangle < 0.68$ eV \cite{heidel}.

The collision energies of the future collider experiments
 would possibly allow for the  study of
the interactions of the particles of
the \rh sector directly.
One interesting reaction which may be possible
  to investigate  in the \nlc is  \cite{eeww}
\be
e^-e^-\rightarrow W^-W^-.
\label{eeww}
\ee
This is  the inverse process to the
neutrinoless double-$ \beta$ decay.
While the reaction $ e^+e^-\rightarrow W^+W^-,$ which will be soon
explored at LEP 200, is possible irrespectively of whether  neutrinos
are Dirac or Majorana particles, the reaction (\ref{eeww}) can
occur  only in the Majorana case and is thus forbidden in
the \smp

The reaction (\ref{eeww}) proceeds via a neutrino exchange in the
t- and u-channels and via a doubly charged triplet Higgs $ \DE^{--}$ exchange
in the s-channel (see Fig. 3). There is a strong cancellation
 between the contributions from the different channels in the amplitude,
which guarantees the  good high energy behaviour of the cross section.
The final state in the process (\ref{eeww}) can be either
$ W_L W_L,$ $ W_R W_R$ or $ W_L W_R.$
 Because the mixing between
$ W_L$ and $ W_R$ is known to be small, the last channel is suppressed.
The  $ W_L W_L$
final state is, in turn,  suppressed by the smallnesses
 of the light neutrino mass
and the triplet Higgs Yukawa coupling to the $ W_L.$
Thus the final state $ W_R W_R $ is the most relevant one, provided that the
collision energy exceeds the kinematical threshold.

\setlength{\unitlength}{4mm}
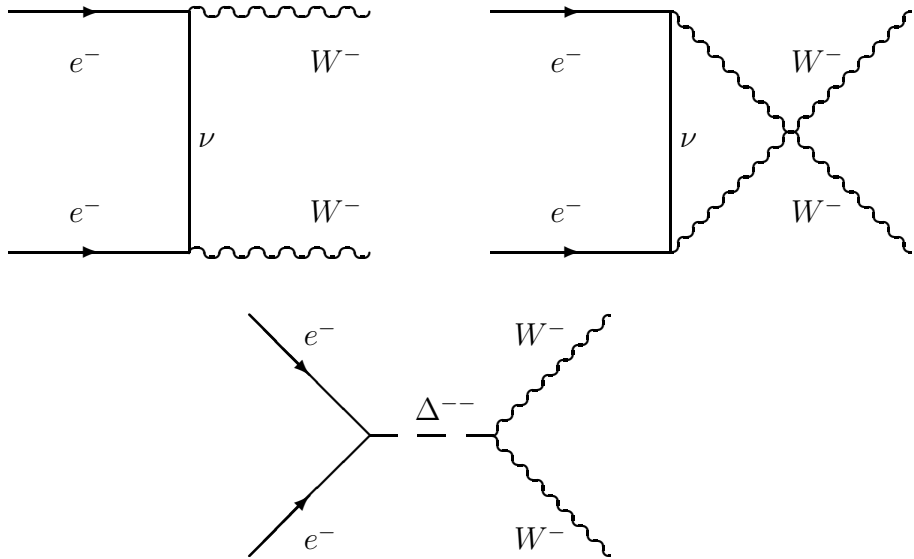
\begin{figure}

\begin{picture}(28,10)(-4,-2)
\thicklines

\put(0,8){\line(1,0){6}}
\put(0,0){\line(1,0){6}}
\put(6,8){\line(0,-1){8}}

\put(0,8){\vector(1,0){3}}
\put(0,0){\vector(1,0){3}}

\multiput(6.25,0)(1,0){6}{\oval(0.5,0.3)[t]}
\multiput(6.75,0)(1,0){6}{\oval(0.5,0.3)[b]}
\multiput(6.25,8)(1,0){6}{\oval(0.5,0.3)[t]}
\multiput(6.75,8)(1,0){6}{\oval(0.5,0.3)[b]}


\put(2,1){\shortstack{\(e^-\)}}
\put(2,6){\shortstack{\(e^-\)}}
\put(10,6){\shortstack{\(W^{-}\)}}
\put(10,1){\shortstack{\(W^{-}\)}}
\put(6.3,3.5){\shortstack{\(\nu \)}}


\put(16,8){\line(1,0){6}}
\put(16,0){\line(1,0){6}}
\put(22,0){\line(0,1){8}}

\put(16,8){\vector(1,0){3}}
\put(16,0){\vector(1,0){3}}

\multiput(22.,0.25)(0.5,0.5){16}{\oval(0.5,0.5)[br]}
\multiput(22.5,0.25)(0.5,0.5){16}{\oval(0.5,0.5)[tl]}

\multiput(22.,7.75)(0.5,-0.5){16}{\oval(0.5,0.5)[tr]}
\multiput(22.5,7.75)(0.5,-0.5){16}{\oval(0.5,0.5)[bl]}


\put(18,1){\shortstack{\(e^-\)}}
\put(18,6){\shortstack{\(e^-\)}}
\put(26,1){\shortstack{\(W^-\)}}
\put(26,6){\shortstack{\(W^-\)}}
\put(22.3,3.5){\shortstack{\(\nu \)}}

\end{picture}

\begin{picture}(20,10)(-12,-2)
\thicklines

\put(0,8){\vector(1,-1){2}}
\put(0,0){\vector(1,1){2}}
\put(4,4){\line(-1,1){4}}
\put(4,4){\line(-1,-1){4}}

\multiput(11.5,7.75)(-0.5,-0.5){8}{\oval(0.5,0.5)[br]}
\multiput(12,7.75)(-0.5,-0.5){8}{\oval(0.5,0.5)[tl]}

\multiput(8.5,3.75)(0.5,-0.5){8}{\oval(0.5,0.5)[bl]}
\multiput(8,3.75)(0.5,-0.5){8}{\oval(0.5,0.5)[tr]}

\multiput(4,4)(1.6,0){3}{\line(1,0){0.9}}


\put(1.8,7){\shortstack{\(e^-\)}}
\put(1.8,0.2){\shortstack{\(e^-\)}}
\put(8.8,7){\shortstack{\(W^-\)}}
\put(8.8,0.2){\shortstack{\(W^-\)}}

\put(5.5,4.5){\shortstack{\(\Delta^{--}\)}}

\end{picture}

\caption{Feynman diagrams for the inverse  neutrinoless
double-$ \beta$ decay.}

\end{figure}

The phenomenological aspects of the reaction (\ref{eeww})
have been previously studied in ref.\cite{eeww}.
In Paper II it is investigated in a
 more general approach by
assuming a general form for the couplings involved and
 taking into account the
polarization of the initial and final state particles.
General conditions which  the couplings should satisfy
to ensure the good high energy behaviour of the process
are derived there.
These conditions concern the case where the final state $ W$'s
are longitudinally polarized, since the possible divergences would occur
in this channel.
In  gauge theories the longitudinal components of massive
gauge bosons play a special role since they are created by the Higgs
mechanism. In the original Lagrangian they correspond to the Goldstone
bosons. At high energies the gauge boson
interactions are indistinguishable from the corresponding
 Goldstone boson interactions.
 This result is known as
the equivalence theorem \cite{equiv}.
Since the singly charged Goldstone  boson interaction with $ \DE^{--}$
follows from the potential  (\ref{tripot}),
 studies of the process (\ref{eeww}) could  give  valuable information
about the triplet Higgs potential.

If  $ W_R$'s were too heavy to be pair produced
in the \nlc, one could still  see  signals of
doubly charged triplet Higgses in  $e^-e^- $ collisions through
other processes.
For example,  one can
look for a s-channel    $\DE^{--}$ resonance  in the reaction
\be
e^-e^-\rightarrow l^- l^-,
\ee
where $ l^-$ can be $ e^-,$ $ \mu^-$ or $ \tau^-.$
 The uncertainty associated with this process is that it depends
on the  unknown $ \DE^{++}l^-l^-$ Yukawa coupling constant $  h.$
Assuming $ h$ to be of the order of the gauge coupling constant
one will be able to see at the \nlc
the doubly charged Higgs bosons  with masses
 as high as
$ M_{ \DE }=10$ TeV \cite{meie5}. Since the mass $ M_{ \DE }$ is set
by the right-handed symmetry breaking scale, it may be
very large.
In the case of  the susy left-right model, however,
where $ \DE^{--}$  should be light \cite{kati},
 the \nlc with $ \sqrt{s}=1$ TeV
should suffice to discover it.

\subsection{ Searches for Supersymmetric Left-Right Model}

The success of supersymmetric models in explaining the theoretical
ambiguities of the \sm has  not been followed
by an experimental discovery of   supersymmetric particles.
Most of  the mass limits for susy particles
are obtained assuming  the minimal supersymmetric
standard model  and  depend on  various assumptions
on the model parameters. The current best limit for the mass
of lightest neutralino, which is supposed to be also the lightest
supersymmetric particle  obtained by ALEPH experiment at LEP
is \cite{susymass}  $ m_{\tilde{\chi^0}}\gsim 18$ GeV.
The mass limits for   charged susy particles, the
charginos and selectrons, obtained
from LEP, are \cite{susymass}
 $ m_{\tilde{\chi}^+,\tilde{e}^+}\gsim
45$ GeV. In these analyses the mass of the left-handed selectron is usually
assumed to be  larger than the mass of the \rh selectron
as follows from the minimal supersymmetric standard model
with  the unification assumption \cite{ibanez3}.
The coloured states are also supposed to be heavier than the uncoloured
states \cite{colour}, and
therefore the squark production in colliders is disfavoured
 compared with the slepton production.
The corresponding mass bounds for the susy left-right model particles  are
expected  to be similar for the charged particles, since the bounds
 are rather model independent and are set mainly by the collision energy of
LEP. The experimental bounds for neutralinos, however, are model dependent
and should be considerably weakened when the  model dependent assumptions
are relaxed.

There is an upper limit of ${\cal O}( 1)$ TeV  for the susy breaking scale
  \cite{haber,colour}, above which the mass differences between particles and
their supersymmetric partners become too large to cancel the
contributions to the  scalar self-masses to a sufficient level.
Since the masses
of susy particles are set by  soft mass terms, they cannot exceed this
limit.
This implies that susy models in the present context
can be discovered or excluded after realization
of the \nlc and the Large Hadron Collider projects.

Tests of the minimal supersymmetric standard model in $ e^-e^-,$ $ e^-\g$ and
$ \g\g$ collisions at the \nlc have been considered in ref.\cite{ruckl}.
In Paper III and Paper IV of this Thesis the possible tests of the susy
left-right model using  the same collision modes have been investigated.

One very promising  test of the susy \lr in the \nlc is
the production of doubly charged higgsinos $ \tilde{\DE}^{++},$ discussed in
Paper III.
They  carry   two units of lepton number and  decay
to two  leptons with equal charge and large missing energy,
giving a  very clean signature in experimental search.
Their mass is a free parameter in the model and, as argued in Section 4,
 should not differ too much from the electroweak symmetry
breaking scale.
In Paper III  we have  studied  the triplet higgsino production
in  $ e^-e^-,$ $ e^+e^-,$ $ e^-\g$ and $ \g\g$ options of the
\nlc via the processes
\be
e^+e^-\rightarrow \tilde{\DE}^{++}\tilde{\DE}^{--},
\label{p1}
\ee
\be
e^-e^-\rightarrow \tilde{\chi}^{0}\tilde{\DE}^{--},
\label{p2}
\ee
 \be
\g e^-\rightarrow \tilde{l}^{+}\tilde{\DE}^{--},
\label{p3}
\ee
\be
\g\g\rightarrow \tilde{\DE}^{++}\tilde{\DE}^{--}.
\label{p4}
\ee
With the collision energy of $ \sqrt{s}=1$ TeV the cross section of the process
(\ref{p1}) is found to be of the order of $ {\cal O}(1)$ pb for the
large range of particle masses. This would
allow for the  discovery of  $\tilde{\DE}^{--} $
with a mass close to the beam energy.  The cross sections of
the processes (\ref{p2}) and
(\ref{p3}) are in general about an order of magnitude smaller,
but since the single $\tilde{\DE}^{--} $ production is associated with
the production of another, presumably lighter, particle, then
the study of the processes  (\ref{p2}) and (\ref{p3})
will enlarge the   $\tilde{\DE}^{--} $ mass range testable in the \nlcp
Despite of the smaller collision energy of  $ \g\g$ compared with  $ e^+e^-$
collisions, the virtue of the reaction (\ref{p4}) is that it
depends   only on the unknown  doubly charged higgsino mass.

If the doubly charged
higgsinos are too heavy to be produced in the future colliders,
 their extra contribution as the virtual intermediate states in
 the selectron pair production,
\be
e^+e^-\rightarrow \tilde{e}^+\tilde{e}^-,
\label{selpr}
\ee
in LEP 200 and  \nlc might reveal their existence. This possibility
has been studied in Paper IV.
Because of the larger number of neutralinos,
compared with the minimal supersymmetric standard model,  and the doubly
charged higgsino involved in the process,
the  cross section of the reaction (\ref{selpr}) is found
in the susy \lr  to be  about $  5$ times larger than in the minimal
supersymmetric standard model.
 The effects  of the doubly
charged higgsino can be found by studying the distribution of  final state
electrons. Since $\tilde{\DE}^{--} $ contribution to the process (\ref{selpr})
compared with the neutralino contribution is with the opposite
angular distribution,
 for  a large range of model parameters the doubly
charged higgsino contribution should be observable.
This result, however, assumes that at least the right
selectrons are light enough to be pair produced in the colliders.

\newpage

\section{Summary}

While the Standard Model of electroweak interactions is known to be in
a good agreement with the data collected so far, it is not excluded that
new phenomena beyond it could start to manifest themselves when
the experimental search moves up to the TeV-scale.
The aim of this Thesis is to study the potential of
the  non-conventional
$ e^-e^-,$ $ e^-\g$ and $ \g\g$ collision  modes of the \nlc
for testing phenomena beyond the \smp
These collision modes, which previously have been available at
much lower energy scales than  the \nlc will offer,
have  several advantages over the
$ e^+e^-$ mode since they allow to study  several  aspects of
the underlying physics which are not accessible  in the conventional
operation mode.

The $ e^-\g$ collision mode is particularly suitable for studying the
three boson coupling $ \g WW$ through the reaction $ e^-\g\rightarrow
W^-\nu.$  In this reaction  the $ \g WW$ coupling can be measured
independently
from the $ ZWW$ coupling, in contrast with, for example, the reaction
$ e^+e^-\rightarrow W^+W^-.$
Assuming Lorentz  and $CP$ invariance and  imposing
 $ U(1)_{ em }$ symmetry,
the general  $ \g WW$ coupling  can be parametrized
with two independent form factors $ \ka$ and $ \la$.
We have analysed the sensitivity of the reaction $ e^-\g\rightarrow
W^-\nu$
 to these parameters,  assuming beam polarization  and
taking into account the anticipated  developments in the \nlc design.
With c.m. energy $ \sqrt{s}=420$ GeV and
data of 50 fb$ ^{-1},$  at  90\% CL the parameters may be constrained to
the range   $ -0.01\le 1-\ka\le 0.01,$ $ -0.012\le\la\le 0.007.$

Motivated by its capability to explain the smallness of neutrino masses
we have studied  the \ssu model, in which  the left-right symmetry
is spontaneously broken by a   right-handed triplet Higgs field.
We have investigated a particularly interesting process, the so-called
inverse double-$ \beta$ decay $ e^-e^-\rightarrow W^-W^-$,
mediated  by the Majorana neutrinos in t- and u-channel and  by the doubly
charged Higgs bosons, which carry lepton number two, in s-channel.
In Paper II we derive helicity amplitudes for this reaction and give the
 conditions for the couplings
to ensure its good high energy behaviour.
The process is
 useful not only for clarifying the nature of neutrinos but also
for studying  the Higgs sector of the theory.

In order   to solve the hierarchy problem
associated with the quadratic self-energy divergences of Higgs bosons
 the \lr is  supersymmetrized and we have studied the consequences
of this model in the experiments at the \nlcp
The most distinctive signature of the model, two same charge leptons
with missing energy,  would be provided by the
decay of the doubly charged higgsino.
Using all the collision modes of the \nlc one will be able to
study doubly charged higgsino
 masses almost up to the c.m. energy of the collider. If the doubly charged
higgsino is too heavy to be produced in  experiments,
it can possibly be discovered
due to    its   contribution to the angular distribution
of the selectron pair production.

The present experimental results give upper bounds for the masses
of the new particles predicted by the \lrp
They do not, however, exclude the possibility considered in this work
that some indications of the left-right symmetry would be  discovered
in the \nlcp
On the other hand, if such evidences were not found, it would not
mean  that the \lr is excluded.
Actually, according to some
SO(10) grand unified scenarios, the \rh symmetry breaking scale
is $ 10^{10-12}$ GeV \cite{soten}, much above the scale achievable in
future accelerators.

\end{document}